\documentclass{svproc}
\usepackage{url}
\usepackage{subcaption}
\usepackage{physics}
\usepackage{graphicx}
\usepackage{amsmath}
\usepackage{amssymb}
\usepackage{amsfonts}
\usepackage{mathtools}
\usepackage{hyperref}

\begin{document}
\mainmatter              %
\title{Chaos, Fractals, and Ergodicity in Nonlinear Quantum Protocols}
\titlerunning{Ergodicity in Nonlinear Quantum Protocols}

\author{Attila Portik\inst{1,2,3}, Orsolya K\'alm\'an\inst{1}, Tam\'as Kiss\inst{1}}
\authorrunning{Attila Portik et al.} %

\institute{HUN-REN Wigner Research Centre for Physics, H-1525   Budapest, Hungary \\
\email{portik.attila@wigner.hun-ren.hu}
\and
ELTE Eötvös Loránd University, H-1117 Budapest, Hungary
\and
Qutility @ Faulhorn Labs, H-1117 Budapest, Hungary
}

\maketitle              %

\begin{abstract}
Iterated nonlinear quantum protocols provide a straightforward way of realizing nonlinear dynamics in quantum systems through measurement and postselection. We investigate a family of protocols in which the order of the induced nonlinearity is controlled by the number of qubits involved in a single step. By examining the resulting rational maps and applying a set of numerical diagnostics, we identify extended parameter regions that exhibit ergodic-like dynamics for protocol orders $n=3,4,$ and $5$. We further characterize the associated chaotic behavior through the distribution of Lyapunov exponents. Our results show that ergodic-like dynamics persists beyond the quadratic case, while the geometry of the corresponding parameter regions and the typical degree of chaoticity depend strongly on the order of the nonlinear transformation.

\end{abstract}

\section{Introduction}

Measurement-induced postselection provides a simple and physically accessible route through which elementary quantum circuits can generate genuinely nonlinear evolution on quantum states. While the linearity of quantum mechanics usually prevents the emergence of classical-like chaos, postselection breaks the deterministic and unitary nature of the system and turns the time evolution of the surviving subsystem into a probabilistic and effectively nonlinear process \cite{aaronson2005quantum,terno1999nonlinear,abrams1998nonlinear}. When such transformations are applied iteratively, with the output of one successful step serving as the input of the next, they naturally define a discrete-time dynamical system on the quantum state space. Even simple protocols built from multi-qubit entangling gates, projective measurements, postselection, and a single parameterized one-qubit unitary already give rise to a remarkably rich class of nonlinear dynamics, including stable and unstable fixed points, competing attractors, fractal basin boundaries, and exponential sensitivity to initial conditions. This establishes iterated nonlinear quantum protocols as a natural bridge between quantum information processing and nonlinear dynamical systems, providing both a physically motivated setting for studying complex dynamical phenomena in quantum systems and a flexible framework for applications in quantum state manipulation, distillation, and validation. 

This connection between iterated quantum protocols and nonlinear dynamics has led to the identification of several characteristic phenomena. Earlier work showed that measurement-induced nonlinear transformations can be used for quantum state matching and state-discrimination tasks \cite{KalmanKissJex2018}. Later studies revealed distinct nonlinear features including competing attractors, fractal basin boundaries, and strong sensitivity to initial conditions \cite{Kiss2006,Kiss2011,Gilyen2016}. For mixed initial states, noisy inputs were shown to induce a purity-driven transition: above a critical purity, the dynamics retains a Julia set-like boundary structure separating asymptotic outcomes, while below it the fractal structure disappears and the dynamics becomes globally regular \cite{Malachov_2019,PORTIK2022127999}. Some of these characteristics of the nonlinear protocols were later demonstrated experimentally in a photonic implementation \cite{qu2021observation}. More recent work showed that the chaotic dynamics remains robust under realistic imperfections, with moderate preparation noise and coherent gate errors deforming but not destroying the underlying fractal structure \cite{robustness2023}. Related nonlinear constructions have also enabled universal and unambiguous concentration and distillation of Bell pairs \cite{Kalman2025}. Together, these results establish iterated nonlinear quantum protocols as both useful quantum-information primitives and possible experimental realizations of complex nonlinear dynamics. Recently, we have further expanded this picture by identifying a class of iterated nonlinear quantum protocols that exhibit globally chaotic, strongly mixing, and ergodic-like dynamics \cite{Portik2026}. For a particular choice of the final single-qubit unitary, the pure-state dynamics explores the entire Bloch sphere, providing a physical realization of ergodic behavior in a nonlinear quantum map. 

\section{The nonlinear protocol}

\begin{figure}[!htbp]

\centering
\includegraphics[width=0.65\linewidth]{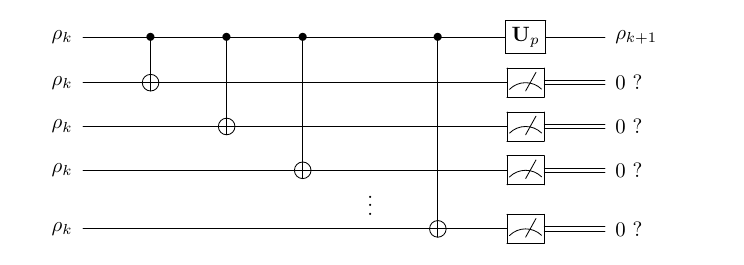}
\caption{Quantum circuit realising the $(k+1)$th step of an $n$th order nonlinear quantum protocol. $n$ identical and independent qubits, all in the initial state $\rho_{k}$ are acted upon by $n-1$ two-qubit controlled-NOT gates, in which the uppermost qubit is the control bit, while the rest of the $n-1$ qubits are the targets. Then, the $n-1$ target qubits are measured in the computational basis, and if all of the measurements result in "$0$", then a parametrized single-qubit gate is applied on the unmeasured qubit, the state of which is transformed into $\rho_{k+1}$.
}
\label{fig:1step}
\end{figure}

In this work, we investigate a family of iterated nonlinear quantum protocols in which the order of the induced nonlinearity is controlled by the number of qubits participating in a single protocol step. Our goal is to understand how the order of nonlinearity influences the chaotic and ergodic-like properties of the resulting dynamics. A single step of the protocol is illustrated in Fig.~\ref{fig:1step}. First, $n$ identically prepared qubits are entangled using a star-shaped network of CNOT gates, with the first qubit acting as the common control. The remaining $n-1$ qubits are then measured and postselected onto the state $\ket{0}$. Conditioned on a successful outcome, the surviving qubit undergoes a parameterized single-qubit unitary transformation $U_p$. Together, these operations induce an effective $n$th-order nonlinear transformation of the qubit state, described by a rational map on the Bloch sphere.

In practice, the nonlinear transformation is not applied only once, but rather in an iterative manner, meaning that the output state of one successful step becomes the input state for the next. Specifically, at the $(k+1)$-th iteration, one prepares $n$ qubits in the same state $\ket{\psi}_k$, where $\ket{\psi}_k$ is obtained by performing $k$ consecutive nonlinear steps starting from the initial state $\ket{\psi}_0$. The same effective nonlinear transformation is then applied as before, yielding the updated state $\ket{\psi}_{k+1}$. Repeating this procedure generates a discrete-time dynamical system on the single-qubit state space. This iterative structure is a defining feature of the protocol. Each step is nonlinear and probabilistic, yet locally simple. However, their repeated application induces a complex dynamical system with exotic characteristics, such as competing attractors and basins of attraction separated by fractal boundaries, ultimately giving rise to chaotic behavior.

\section{Nonlinear Time Evolution}

Since we consider an ideal implementation without state-preparation noise or imperfections during the protocol, the qubits remain in pure states throughout the evolution. We can therefore use state vectors to track the transformation of the qubit states. Consequently, two real parameters, or equivalently a single complex parameter, are sufficient to describe the state of a qubit. Using the stereographic parametrization of the Bloch sphere, an arbitrary pure state can be written as
\begin{equation}
\ket{\psi(z)}
=
\frac{\ket{0}+z\ket{1}}{\sqrt{1+|z|^2}},
\qquad
z\in\hat{\mathbb{C}},
\label{eq:pure_state_parametrization}
\end{equation}
where $\hat{\mathbb{C}}=\mathbb{C}\cup\{\infty\}$ is the extended complex plane. The points $z=0$ and $z=\infty$ correspond to the computational-basis states $\ket{0}$ and $\ket{1}$, respectively.
A single step of the protocol acts on $n$ identically prepared qubits. The CNOT network followed by postselection of all measured qubits onto $\ket{0}$ induces the nonlinear transformation
\begin{equation}
z \longmapsto z^n,
\label{eq:power_map}
\end{equation}
with success probability
\begin{equation}
P_{\mathrm{succ}}(z)
=
\frac{1+|z|^{2n}}
{\left(1+|z|^2\right)^n}.
\label{eq:success_probability}
\end{equation}

After a successful postselection the single-qubit unitary
\begin{equation}
U_p
=
\frac{1}{\sqrt{1+|p|^2}}
\begin{pmatrix}
1 & p\\
-\bar p & 1
\end{pmatrix},
\qquad
p\in\mathbb C ,
\label{eq:Up}
\end{equation}
is applied on the surviving qubit. Combining the nonlinear transformation with the unitary operation yields the rational map
\begin{equation}
f_p(z)
=
\frac{z^n-\bar p}
{1+pz^n}.
\label{eq:nonlinear_complex_map}
\end{equation}

The protocol is applied iteratively, such that the output state of one successful step becomes the input state of the next step. Consequently, the dynamics of pure states is described by
\begin{equation}
z_{k+1}=f_p(z_k),
\qquad
z_k=f_p^{\circ k}(z_0),
\label{eq:iterated_map}
\end{equation}
where $f_p^{\circ k}$ denotes the $k$-fold composition of the map. The conditional quantum evolution is therefore equivalent to the iteration of a degree-$n$ rational map on the Riemann sphere, allowing the protocol to be analyzed using the methods of complex dynamical systems.

\section{Ergodic dynamics in iterated quantum protocols}
\label{sec:ergodic_criteria}

The quadratic member of the family introduced above, $f_p(z) = \frac{z^2-\bar{p}}{1+pz^2},$ was investigated in detail in our previous work \cite{Portik2026}. A particularly important parameter value is $ p=i $, for which the given quadratic map is a Latt\`es map and therefore represents a rare analytically understood example of globally chaotic rational dynamics \cite{Lattes,Milnorbook}. Its Fatou set is empty, while its Julia set coincides with the entire Riemann sphere. Consequently, there are no attracting periodic orbits, and almost every pure initial state exhibits chaotic evolution. Moreover, the dynamics is ergodic and strongly mixing with respect to its invariant measure, meaning that long individual trajectories explore the entire state space, i.e. the Bloch sphere, while ensembles initially localized in a small region are stretched and spanned to the whole surface \cite{Halmos,Viana,Moore}.

The Latt\`es point $p=i$ is, however, not the only parameter value for which the quadratic protocol displays such behavior. We numerically examined the complex parameter plane and identified an extended region around the Latt\`es point in which the dynamics exhibit several characteristic signatures of ergodicity \cite{Gilyen2016,Portik2026}. Since a rigorous proof of ergodicity is generally unavailable for these non-Latt\`es maps, we introduced four distinct and complementary diagnostics, each probing a different aspect of the dynamics and serving as a numerical hallmark of ergodic behavior. We evaluated these criteria across the parameter space and classified as ergodic-like those parameter values for which all four numerical conditions described below are simultaneously satisfied.

The first diagnostic is the absence of attracting periodic orbits. Since every attracting cycle of a rational map attracts at least one critical point \cite{Milnorbook}, it is sufficient to examine the critical orbits. The maps $f_p$ possess only two critical points, $z=0$ and $z=\infty$, each of multiplicity $n-1$. For every parameter value, both critical points were iterated for $10^7$ steps and tested for convergence to periodic cycles of length up to $500$. Parameters for which either critical orbit approached an attracting cycle were excluded.

The second diagnostic is based on the invariant visitation distribution generated by long trajectories. For ergodic-like dynamics, sufficiently long orbits should densely explore the accessible part of the Bloch sphere and produce a stable visitation histogram. Time-averaged distributions were obtained from $100$ random initial states evolved for $10^7$ iterations and recorded on a $500\times500$ equal-area discretization of the sphere.

The third diagnostic examines ensemble spreading. Initial states sampled from a small spherical patch should rapidly disperse and cover the entire Bloch sphere under strongly mixing dynamics. We therefore evolved ensembles of $10^7$ states from $100$ randomly selected patches of area $10^{-4}$ for $100$ iterations and compared the resulting ensemble-averaged distributions with the corresponding time-averaged distributions. Agreement between the two provides numerical evidence for the equivalence of time and ensemble averages.

The fourth diagnostic is the spherical Lyapunov exponent,
\begin{equation}
\lambda(z_0)
=
\lim_{K\rightarrow\infty}
\frac{1}{K}
\sum_{k=0}^{K-1}
\ln
\left[
n|z_k|^{n-1}
\frac{1+|z_k|^2}
{1+|z_k|^{2n}}
\right],
\label{eq:lyapunov_explicit}
\end{equation}
which measures the average exponential separation of nearby trajectories. To compare different protocol orders, we use the normalized exponent
\begin{equation}
\lambda_{\mathrm{norm}}
=
\frac{2\lambda}{\ln n},
\label{eq:normalized_lyapunov}
\end{equation}
where $\ln(n)/2$ corresponds to the maximal-entropy Latt\`es limit. Finite-time exponents were estimated from trajectories of length $10^7$ for $100$ randomly chosen initial states. We required the exponents to be positive and approximately independent of the initial condition.

In the quadratic case, the simultaneous application of these diagnostics revealed an extended connected region of ergodic-like parameters \cite{Portik2026}. Here, we apply the same framework to the higher-order maps and classify a parameter value as ergodic-like only when all four diagnostics are satisfied simultaneously.
\section{Ergodic-like Behavior in Higher-Order Protocols}

\begin{figure}[ht]
\centering

\includegraphics[width=0.35\linewidth]{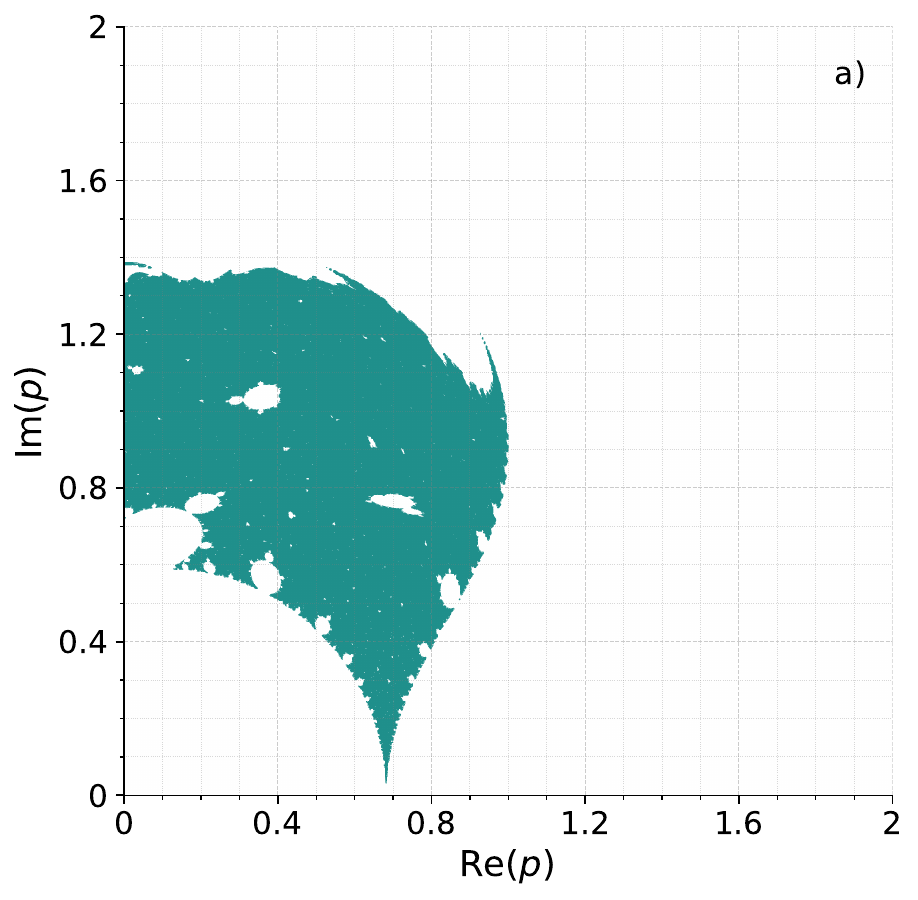}
\includegraphics[width=0.35\linewidth]{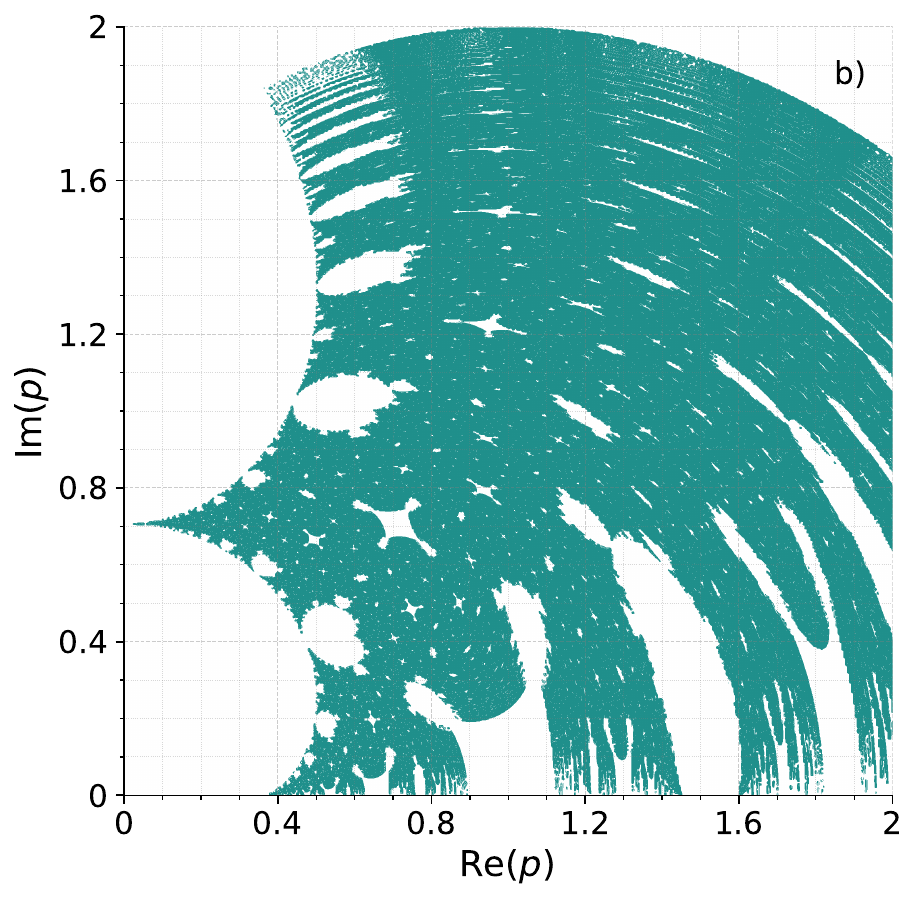}

\includegraphics[width=0.35\linewidth]{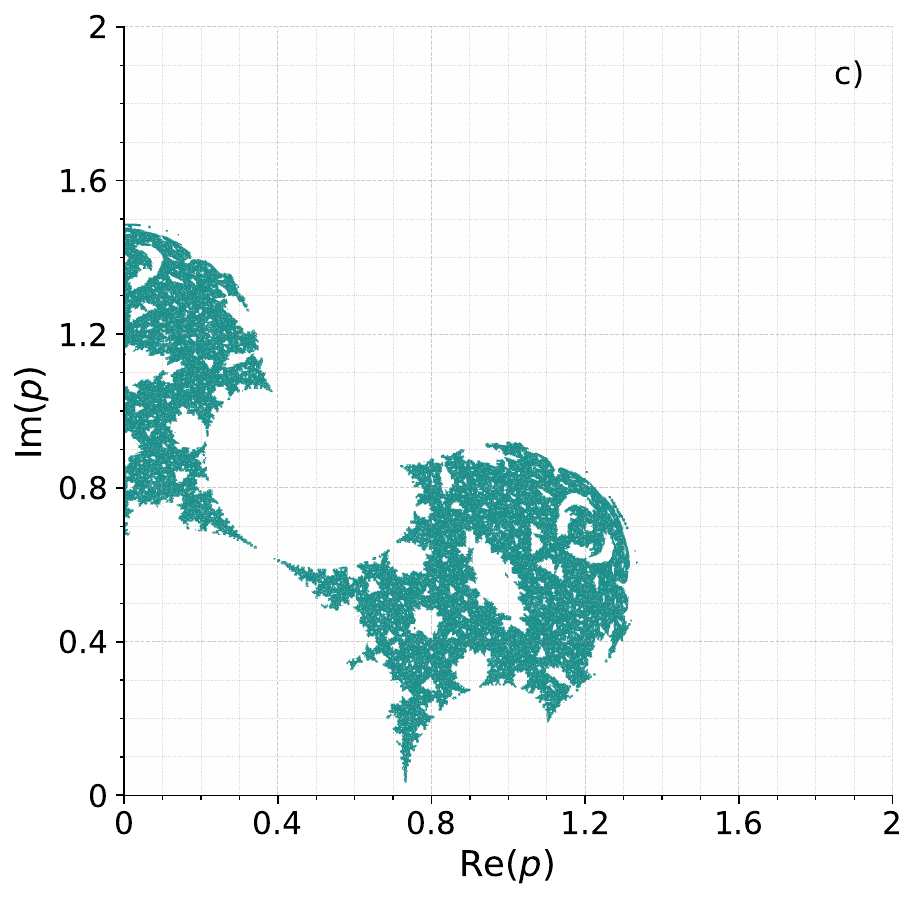}
\includegraphics[width=0.35\linewidth]{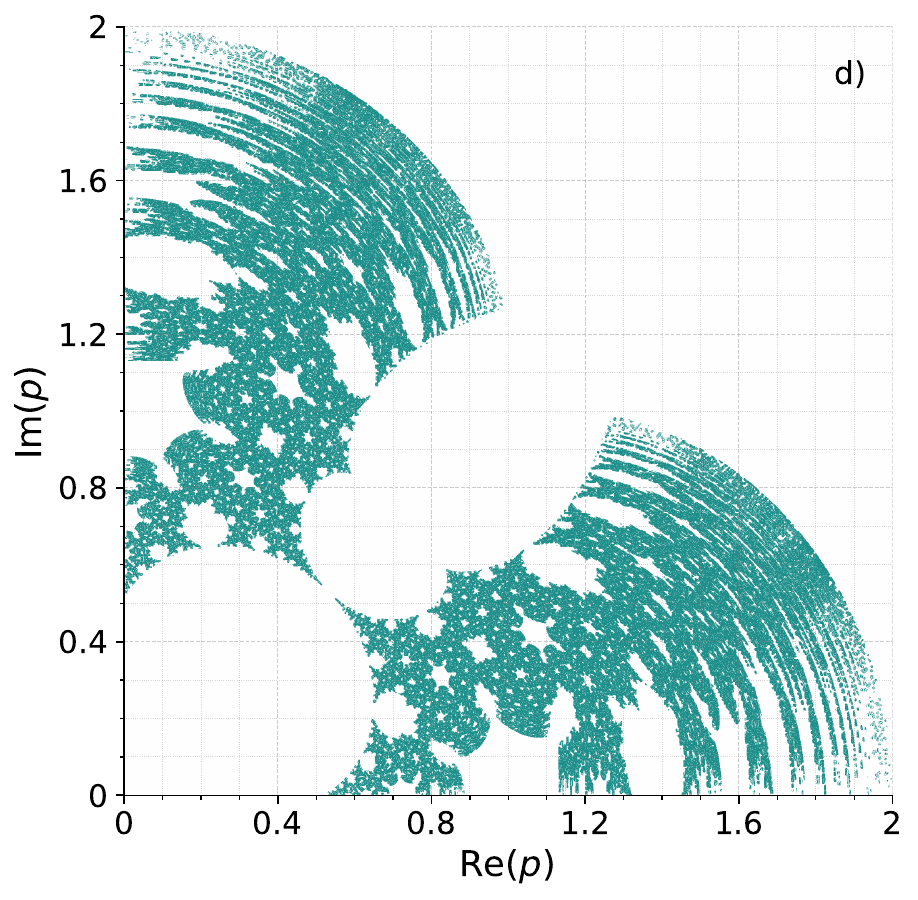}

\caption{
Comparison of the ergodic-like parameter regions of the generalized nonlinear quantum protocol for orders $n=2,3,4,$ and $5$. The displayed regions correspond to parameter values for which all four ergodicity criteria are satisfied simultaneously. The same parameter window, $\mathrm{Re}(p),\mathrm{Im}(p)\in[0,2]$, and numerical resolution, $2000 \times 2000$,  were used in all cases.
}
\label{fig:ergodic_regions}
\end{figure}

We now apply the four ergodicity diagnostics introduced above to the higher-order nonlinear quantum protocols corresponding to $n=3,4,$ and $5$. As in the quadratic case, the diagnostics were evaluated throughout the complex parameter plane, and a parameter value was classified as ergodic-like only if all four criteria were satisfied simultaneously. The resulting ergodic-like parameter regions are shown in Fig.~\ref{fig:ergodic_regions}. For all orders, we find connected region of ergodic-like behavior, demonstrating that the globally chaotic dynamics previously observed in the quadratic protocol persists for higher-order nonlinearities as well. The geometry of these regions, however, changes substantially with the order of the protocol. In addition to the emergence of increasingly intricate structures, the locations of the ergodic-like domains shift within the parameter plane.
\vspace{-2.5ex}

The parameter regions also exhibit the discrete rotational symmetry of the underlying nonlinear map. Since parameters related by
\begin{equation}
p \longrightarrow p\,e^{2\pi i k/(n-1)},
\qquad
k=0,1,\ldots,n-2,
\end{equation}
generate dynamically equivalent systems, the corresponding ergodic-like regions are likewise related by the same rotation. Consequently, it is sufficient to display and analyze a single symmetry sector.

An important difference from the quadratic case is that the special parameter value $p=i$ loses its distinguished role. While for $n=2$ this point corresponds to a Latt\`es map and lies within the ergodic-like region, for higher orders it is not a Latt\`es map. For odd orders it is not even an ergodic-like parameter. This demonstrates that the higher-order protocols exhibit qualitatively different parameter-space structures despite retaining large domains of chaotic behavior.

\begin{figure}[ht]
\centering

\includegraphics[width=0.4\linewidth]{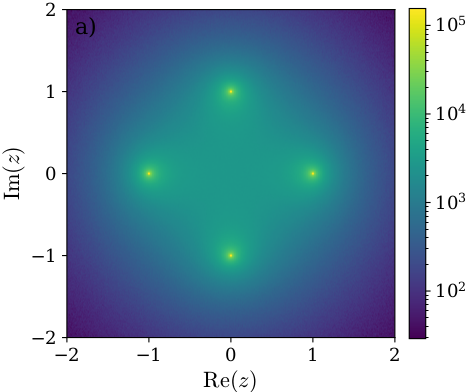}
\includegraphics[width=0.4\linewidth]{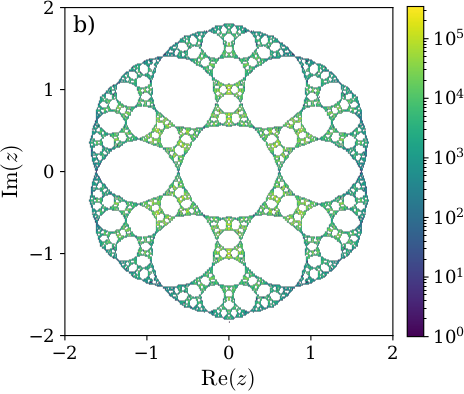}

\includegraphics[width=0.4\linewidth]{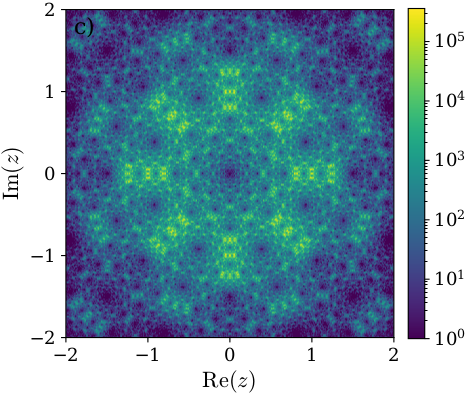}
\includegraphics[width=0.4\linewidth]{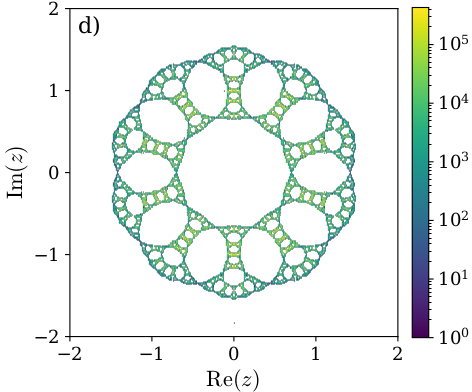}

\caption{
Inverse-iteration densities obtained at the parameter value $p=i$ for protocol orders (a) $n=2$, (b) $n=3$, (c) $n=4$, and (d) $n=5$. In each case, $10^8$ preimages were generated by repeatedly selecting one of the inverse branches with equal probability. The resulting points were accumulated into a $500\times500$ histogram in the complex plane and are displayed using a logarithmic color scale. White regions indicate histogram cells that were not visited during the simulation. \vspace{-3.5ex}
}
\label{fig:lattes_measure}
\end{figure}

\vspace{-2.5ex}

Fig.~\ref{fig:lattes_measure} shows the inverse-iteration densities obtained at $p=i$ to further illustrate this observation. The densities were generated by repeatedly selecting one of the inverse branches of the map with equal probability and recording the resulting backward orbit. Since the preimages of a generic point become dense in the Julia set \cite{Milnorbook}, this procedure provides an efficient numerical method for sampling its invariant measure. While the quadratic protocol generates a smooth density supported on the entire sphere, the higher-order cases exhibit qualitatively different invariant measures. For $n=3$ and $n=5$, the support collapses onto fractal subsets of the state space, the Julia set itself, producing large unvisited regions. In contrast, the case of the $4$th order  protocol remains supported on the entire sphere, but the resulting density is strongly inhomogeneous and exhibits pronounced self-similar structure. These observations demonstrate directly that the special Latt\`es dynamics observed for $n=2$ does not persist for higher-order nonlinearities.

\begin{figure}[ht]
\centering

\includegraphics[width=0.5\linewidth]{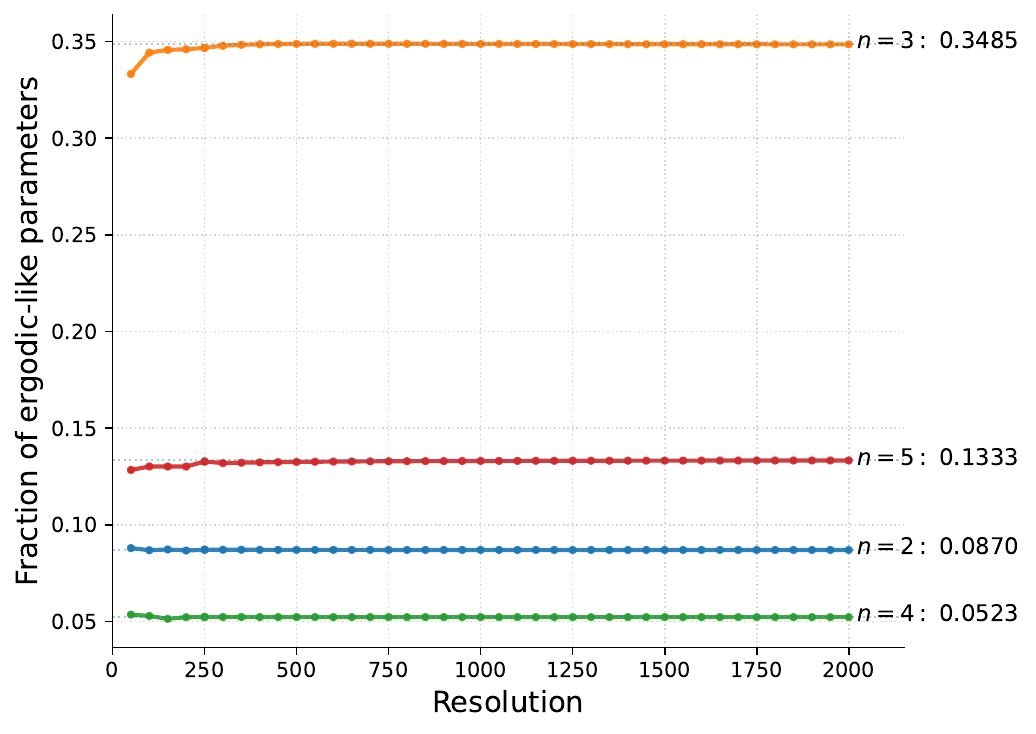}

\caption{
Estimated fraction of ergodic-like parameters as a function of the numerical resolution for protocol orders $n=2,3,4,$ and $5$. The fractions were computed within the fixed parameter window $\mathrm{Re}(p),\mathrm{Im}(p)\in[0,3]$ for increasing resolutions. Solid lines show the convergence towards the infinite-resolution limit, while the labels on the right indicate the corresponding asymptotic estimates. The results demonstrate rapid convergence with increasing resolution and reveal a strong dependence of the relative area occupied by the ergodic-like parameter regions on the degree of nonlinearity.
\vspace{-2.5ex}}
\label{fig:ration}
\end{figure}

We also estimated the fraction of ergodic-like parameters within the parameter space. For this purpose, we considered the larger interval of parameters
$\mathrm{Re}(p),\mathrm{Im}(p)\in[0,3]$,
which, according to our numerical findings, contains all ergodic-like parameters within the first quadrant. The remaining quadrants follow from the rotational symmetry discussed above. For increasing resolutions of the parameter grid, we determined the fraction of parameter values satisfying all four ergodicity criteria. The results are shown in Fig.~\ref{fig:ration}.
The estimated fractions converge rapidly with increasing resolution, indicating that finite-resolution effects are small. The estimated values reveal a strong dependence on the order of the protocol. Among the investigated cases, the cubic protocol exhibits the largest fraction of ergodic-like parameters, occupying approximately $35\%$ of the investigated parameter window, while the quadratic, quartic, and quintic protocols occupy approximately $8.7\%$, $5.2\%$, and $13.3\%$, respectively.

\section{Lyapunov Exponents}

While all four criteria are primarily defined to identify ergodic-like parameter values, the Lyapunov exponent provides additional information about the nature of the resulting dynamics. In addition to distinguishing chaotic from non-chaotic behavior through its sign, its magnitude quantifies the average rate at which initially nearby trajectories diverge. The normalized Lyapunov exponent therefore characterizes the strength of the sensitivity to initial conditions and provides a quantitative measure of the chaotic behavior exhibited by a given protocol. In this section, we investigate the Lyapunov exponents associated with the ergodic-like parameter regions identified above.

\begin{figure}[ht]
\centering

\includegraphics[width=0.45\linewidth]{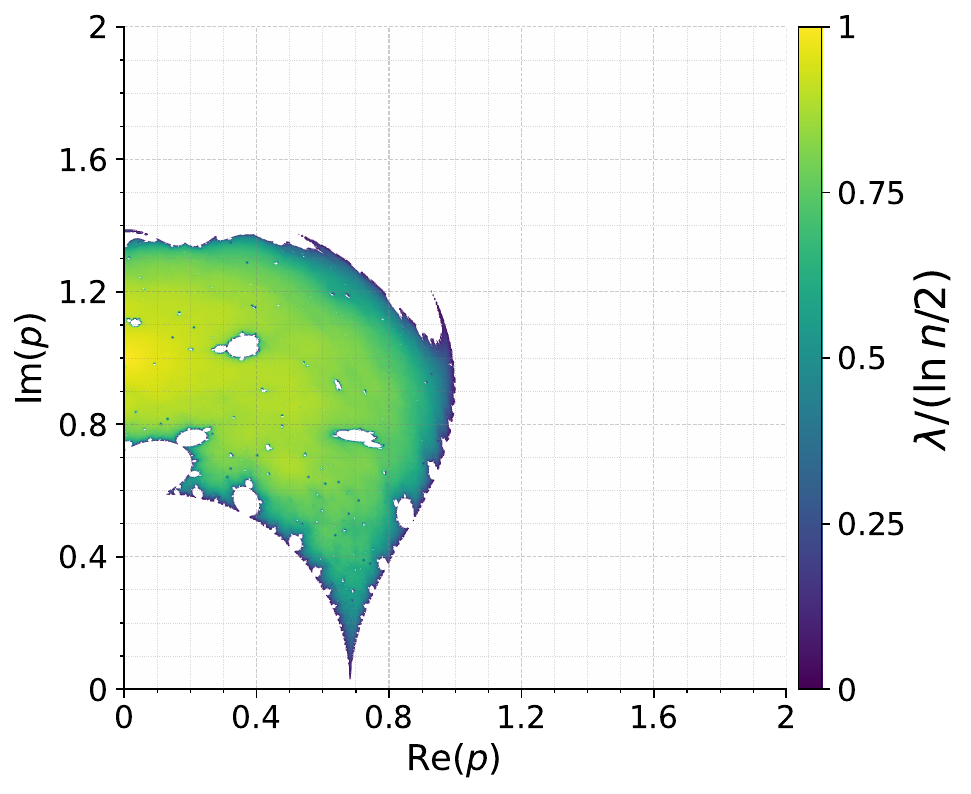}
\includegraphics[width=0.45\linewidth]{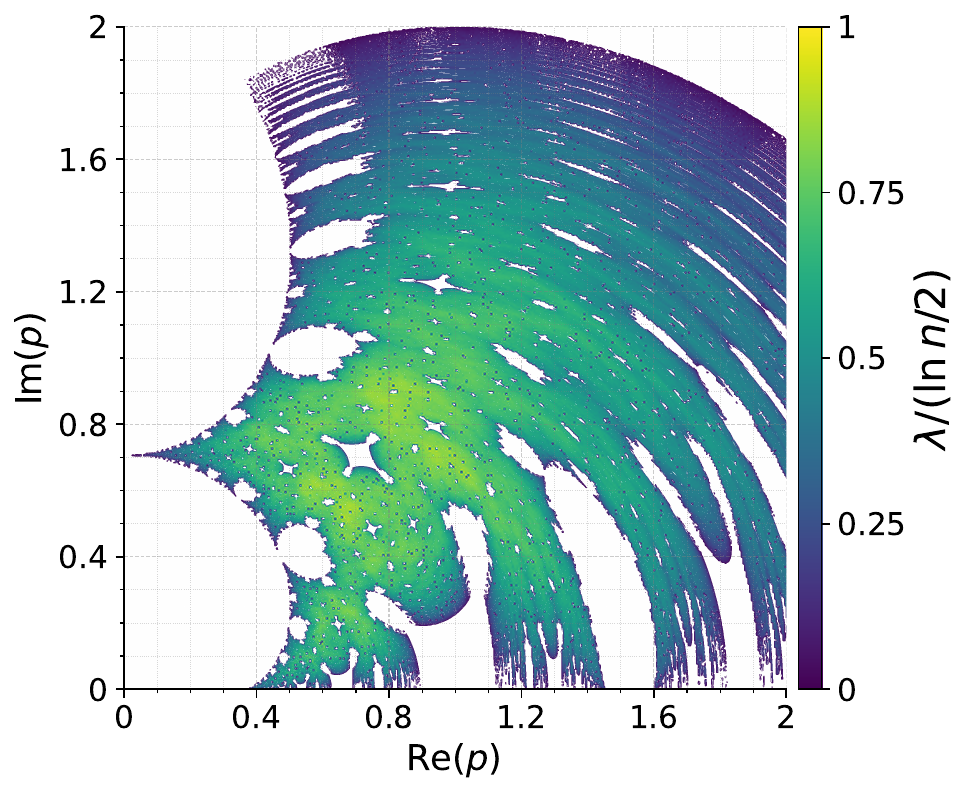}

\includegraphics[width=0.45\linewidth]{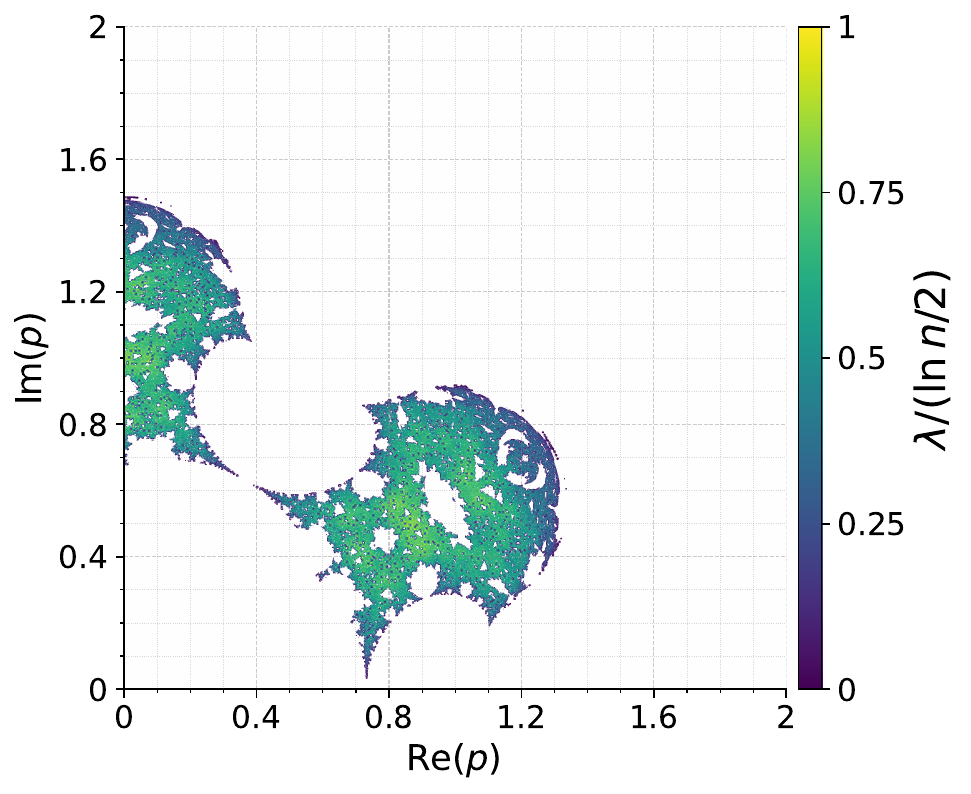}
\includegraphics[width=0.45\linewidth]{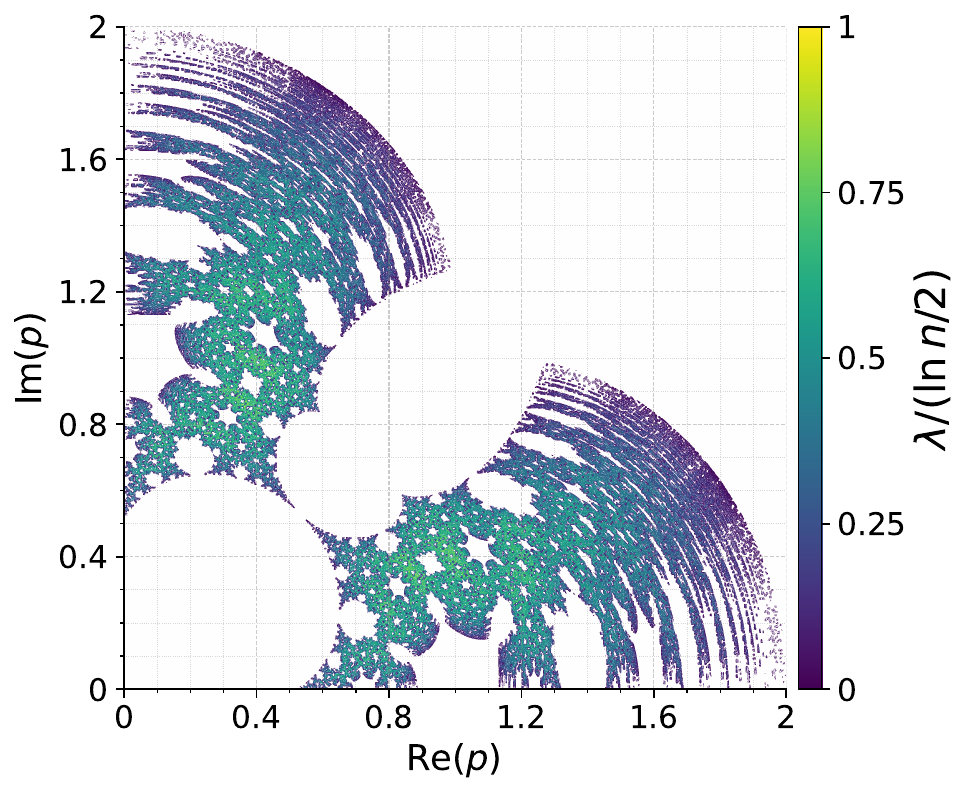}

\caption{
Normalized Lyapunov exponent nonlinear map describing the generalized quantum protocol for protocol orders $n=2,3,4,$ and $5$. The color scale represents the normalized Lyapunov exponent $\lambda/(\ln n/2)$, allowing a direct comparison between different orders. White regions correspond to parameter values for which no ergodic-like behavior was found. While the ergodic-like parameter regions exhibit significantly different geometries for different orders, the normalized Lyapunov exponents remain bounded by unity and display a characteristic decrease towards the boundaries of the ergodic-like regions. All panels were computed on the parameter domain $\mathrm{Re}(p),\mathrm{Im}(p)\in[0,2]$ using a numerical resolution of $2000\times2000$.
}
\label{fig:lyapunov_exponent}
\end{figure}

The normalized Lyapunov exponents obtained for the ergodic-like parameter values are shown in Fig.~\ref{fig:lyapunov_exponent}. For all investigated protocol orders, variations of the estimated exponent are observed within each ergodic-like domain. In general, the largest exponents are found in the interior of the regions, while smaller values occur near their boundaries. This behavior is consistent with the expectation that parameter values located close to transitions towards regular dynamics exhibit weaker instability than those deep inside the chaotic domains.

The quadratic protocol displays the largest normalized Lyapunov exponents, with an extended region approaching the reference value $\lambda_{\mathrm{norm}}=1$, corresponding to the Lattès limit. For higher-order protocols, the exponents are typically smaller and exhibit a considerably more heterogeneous spatial structure. In particular, the cubic and quintic protocols contain broad regions with intermediate Lyapunov exponents, while the quartic case is dominated by values clustered around $\lambda_{\mathrm{norm}}\approx0.6$. These observations indicate that increasing the order of the nonlinear transformation modifies not only the geometry of the ergodic-like parameter regions, but also the strength of the chaotic dynamics realized within them.

\begin{figure}[ht]
\centering

\includegraphics[width=0.5\linewidth]{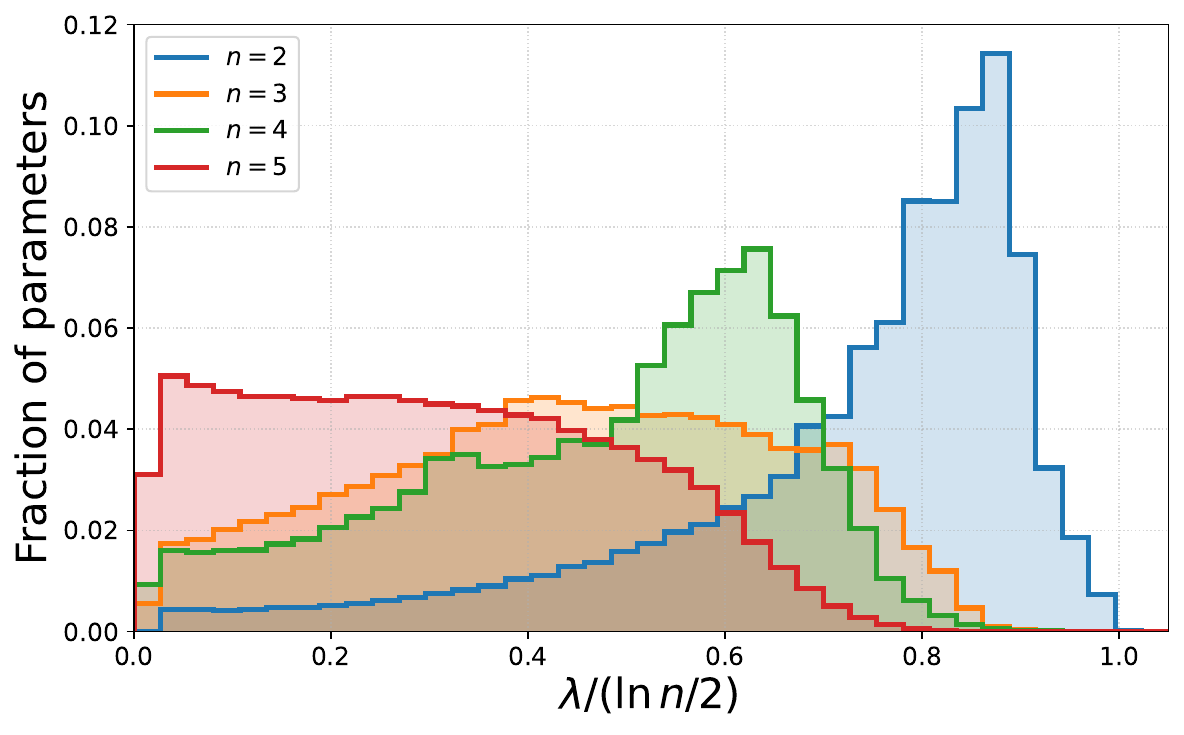}

\caption{
The distribution of the normalized Lyapunov exponent $\lambda/(\ln n/2)$ within the ergodic-like parameter regions for orders $n=2,3,4,$ and $5$. The vertical axis shows the fraction of ergodic-like parameter values contained in each bin. While the maximal normalized Lyapunov exponent remains close to unity for all orders, the distributions exhibit markedly different shapes, indicating substantial variations in the typical degree of chaoticity across the parameter space.
}
\label{fig:hist}
\end{figure}

To characterize the typical Lyapunov exponents associated with each protocol order, we also computed their distributions shown in Fig.~\ref{fig:hist}. The distributions reveal substantial differences between different orders. The quadratic protocol is strongly concentrated at large values, with the majority of ergodic-like parameters yielding normalized exponents between approximately $0.7$ and $0.95$. In contrast, the distributions for $n=3$ and $n=4$ are centered around intermediate values and are noticeably broader. The protocol for $n=5$ exhibits the widest distribution, extending from very small positive exponents to values comparable to those found in the lower-order cases. This broad spread indicates that the corresponding ergodic-like parameter region contains a particularly diverse range of chaotic behaviors. 

The distributions further confirm the trends already suggested by the parameter-space maps. Regions located close to the boundaries of the ergodic-like domains typically exhibit smaller Lyapunov exponents, reflecting the weaker instability of trajectories near transitions to regular dynamics. Consequently, protocols whose ergodic-like parameter regions contain a large boundary-to-area ratio, or consist of intricate structures, exhibit significantly broader and flatter Lyapunov exponent distributions. In contrast, compact ergodic-like regions, such as those observed for the quadratic protocol, are dominated by larger exponents and therefore produce more sharply peaked distributions.

Taken together, these results demonstrate that while ergodic-like dynamics persists for all investigated protocol orders, the strength of the resulting chaos depends strongly on the degree of the nonlinear transformation. The Lyapunov exponent therefore provides information that is not captured by the other diagnostics alone and reveals significant qualitative differences between the chaotic dynamics generated by different higher-order protocols.

\section{Conclusion}

In this work, we extended our previously developed numerical framework for identifying ergodic-like nonlinear quantum protocols to a broader family of higher-order iterated nonlinear quantum protocols. Applying the same set of criteria, we identified finite regions of parameter space for protocol orders $n=3,4,$ and $5$ that exhibit the characteristic signatures of ergodic-like dynamics. These results demonstrate that the emergence of ergodic-like behavior is not restricted to the quadratic protocol but persists for higher-order nonlinearities as well. In addition, we analyzed the magnitudes and distributions of the Lyapunov exponents associated with these parameter regions. Since positive Lyapunov exponents are closely related to the exponential instability and efficient mixing typically observed in chaotic dynamical systems, they provide complementary information about the nature and strength of the chaotic behavior. Our results reveal significant differences in the typical Lyapunov exponents realized by different protocol orders and suggest a connection between the distribution of Lyapunov exponents and the geometric structure of the corresponding ergodic-like regions in parameter space.

\bibliography{references}

\end{document}